# The Effect of Ni and Zn Doping in Bi-2212 from Tunneling Measurements. The MCS Model of the High-$T_c$ Superconductivity in Hole-Doped Cuprates.

A. Mourachkine

*Université Libre de Bruxelles, Service de Physique des Solides, CP233, Boulevard du Triomphe, B-1050 Brussels, Belgium*

We use Ni and Zn impurities doped into $Bi_2Sr_2CaCu_2O_{8+x}$ (Bi2212) to test the MS (Magnetic polaron - Spinon) model proposed earlier. We present electron-tunneling spectroscopy of pure, Ni- and Zn-doped Bi2212 single crystals below $T_c$ using a break-junction technique. We show that magnetic (Ni) and nonmagnetic (Zn) impurities doped into $CuO_2$ planes affect both the $T_c$ and the density of states in Bi2212 but do it differently. In order to explain the data, we had to modify the MS model. Thus, we present a Magnetic Coupling of Stripes (MCS) model. In the MCS model, there is only one superconductivity – a spinon superconductivity along charged stripes. The coherent state of the spinon superconductivity is established by magnons which are excited by motions of charged stripes. So, in the MCS model, the superconductivity has the two mechanisms: along charged stripes and perpendicular to stripes.

## 1. INTRODUCTION

Since the discovery of the superconductivity (SC) in cuprates by Bednorz and Müller [1] in 1986 many theoretical models have been proposed. Recently, Emery with co-workers [2] have presented a model of high-$T_c$ superconductivity (HTSC), which is based on experimental results. The essence of the model is a spinon SC. Spinons are neutral fermions which occur in one-dimensional (1D) physics. The spinon SC occurs on charged stripes in $CuO_2$ planes [3,4]. Stripes are separated by antifferomagnetic (AF) isolating domains. Spinons create pairs on stripes because of the existence of a large spin-gap due to AF correlations [2]. The coherent state of the spinon SC is established at $T_c$ by the Josephson coupling between charged stripes carrying the spinon SC.

In our previous work, we have presented some indications of the coexistence of the spinon SC and predominant $d_{x^2-y^2}$ (hereafter, d-wave) SC in Bi2212 [5]. On the basis of experimental data, we have proposed a MS (Magnetic polaron - Spinon) model of the SC in hole-doped cuprates [5]. In framework of the MS model, the coherent state of the spinon SC is established at $T_c$ via the d-wave polaron SC with the magnetic pairing due to spin-waves (magnons), which occurs in AF domains. A magnetic polaron can be pictured as the electron and a spin polarization around it [6]. The $T_c$ is the characteristic temperature of the d-wave SC. In Bi2212, the $2\Delta/k_B T_c$ value for the d-wave SC with the magnetic pairing lies in a strong coupling range between 5.3 and 5.7 [5]. The magnitude of the spinon-SC gap and the $T_c$ value do not relate to each other. The order parameter (OP) of the spinon SC has either a s-wave or (s+d) mixed symmetry [2,5]. It is important to note that the Josephson current is a sign of the polaron SC since polarons carry charge.

Figure 1 shows an idealized phase diagram for Bi2212 [5]. The maximum magnitude of the spinon-SC gap depends linearly on hole concentration in $CuO_2$ planes, $p$. The dependence of the magnitude of the d-wave SC gap on hole concentration is parabolic since $T_c = T_{c, max}[1 - 82.6(p - 0.16)^2]$, where $T_{c, max} = 95$ K. The maximum magnitude of the spinon-SC gap is larger than the maximum magnitude of the d-wave gap which, however, is more intense than the spinon-SC gap. It is not clear yet what happens with the spinon SC on charged stripes below $p = 0.05$ and above $p = 0.27$. Does it still exist or there is a collapse of the spinon SC at these points. The inset in Fig. 1 shows schematically shapes of the spinon and d-wave SC gaps on the Fermi surface. In addition to main peaks in tunneling spectra, there is a sub-gap. At this moment, it is not clear yet about the origin of this sub-gap. The d-wave SC mediated by magnons can coexist with a g-wave SC which has the same pairing mechanism [7]. We have associated this sub-gap with the g-wave gap. However, more research is needed to find out the exact origin of this sub-gap.

The effect of magnetic and nonmagnetic impurities on the properties of a superconductor provides useful information about the mechanism of the SC. In our case, it is a test of the MS model. In framework of the MS model, we have formulated the Anderson's theorem for cuprates [5]: any impurity doped into $CuO_2$ planes affects both the $T_c$ and the density of states (DOS). Since the tunneling spectroscopy performs direct measurements of the DOS, we have to observe effects caused by impurities. Ni and Zn dopants substitute Cu in $CuO_2$ planes. A Ni is the magnetic impurity. In general, it can be in two oxidation states: $Ni^{2+}(S=1)$ and $Ni^{3+}(S=1/2)$. A Zn is always the nonmagnetic impurity in $CuO_2$ planes, $Zn^{2+}(S=0)$. Ideally, we have to observe this difference between Ni and Zn. Both Ni and Zn will pin stripes in $CuO_2$ planes [3]. A $Ni^{3+}$ oxidation state is formed more easily than $Cu^{3+}$, it should create a weak attractive potential for holes. A large negative potential is expected for Zn impurities, since an inert $d$ shell strongly repels holes, *i.e.* attracts electrons [8]. It is noteworthy that our Ni- and Zn-doped Bi2212 single crystals have similar values of $T_c$, so, this circumstance simplifies the interpretation of our experimental data. In fact, the magnetic origin of the d-wave SC was determined due to measurements in Ni-doped samples [5], hence, direct measurements of the DOS in Ni- and Zn-doped Bi2212 single crystals are important from many points of view and it is a good test for the MS model.

The structure of the paper is as follows. Experimental details are described in Section 2. In Section 3, we consider statistics of gap measurements at low temperature on a pure, a Ni- and Zn-doped Bi2212 single crystals. In Sections 4 and 5, we focus attention on measurements on Ni-doped Bi2212 single crystals. Measurements on Zn-doped Bi2212 samples are presented in Section 6. We discuss all data and the MCS model in Section 7. The final conclusions are presented in Section 8.

## 2. EXPERIMENTAL

The single crystals of Bi2212 were grown using a self-flux method and then mechanically separated from the flux in $Al_2O_3$ or $ZrO_2$ crucibles [9]. The dimensions of the samples are typically $3\times1\times0.1$ mm$^3$. The chemical composition of the Bi-2212 phase corresponds to the formula $Bi_2Sr_{1.9}CaCu_{1.8}O_{8+x}$ in overdoped crystals as measured by energy dispersive X-ray fluorescence (EDAX). The crystallographic $a, b, c$ values of the overdoped single crystals are of 5.41 Å, 5.50 Å and 30.81 Å, respectively. The $T_c$ value was determined by either dc-magnetization or by four-contacts method yielding $T_c$ = 87 - 90 K with the transition width $\Delta T_c \sim 1$ K. Some overdoped single crystals were carefully checked out to ensure that they are in an overdoped phase: the $T_c$ value was increasing up to 95 K when some oxygen was chemically taken off the samples.

The single crystals of Ni- and Zn-doped Bi2212 were grown using also the self-flux method. The chemical composition of the Bi2212 phase with $T_c$ = 75 - 76 K corresponds to the formula $Bi_2Sr_{1.95}Ca_{0.95}(CuNi)_{2.05}O_{8+x}$ and $Bi_2Sr_{1.98}Ca_{0.83}(CuZn)_2O_{8+x}$ as measured by EDAX. The content of Ni is about 1.5 % with respect to Cu and the Zn content is about 1%.

Experimental details of our break-junction (B-J) technique can be found elsewhere [10]. Here we present a short description of some technical details. Many break junctions were prepared by gluing a sample with epoxy on a flexible insulating substrate and then were broken by bending the substrate with a differential screw at 14 - 18 K in a helium atmosphere. By changing the distance between two pieces of a single crystal by a differential screw, it is possible to obtain a few tunneling spectra in one B-J. The normal resistance ($R_N$) of break junctions outside of the gap ranged from 50 Ω to 50 MΩ. The tunneling current-voltage characteristics $I(V)$ and the conductance curves $dI/dV(V)$ were determined by the four-terminal method using a standard lock-in modulation technique. The electrical contacts (typically with a resistance of a few Ω) were made by attaching gold wires with silver paint. The sample resistance (with the contacts) at room temperature varied from 10 Ω to about 2 kΩ, depending on the sample.

## 3. STATISTICS OF THE MAGNITUDE OF TUNNELING GAP

The hole concentration in $CuO_2$ planes of the Bi2212 single crystals with $T_c$ = 87 - 90 K (overdoped) and $T_c$ = 75 - 76 K (Ni- and Zn-doped) calculated from $T_c/T_{c,\ max}$ = 1 - 82.6($p$ - 0.16)$^2$, where $T_{c,\ max}$ = 95 K, is equal to $p$ = 0.19 and $p$ = 0.11, respectively. The maximum magnitude of the gaps for these two cases are shown in Fig. 1. The magnitude of the SC gap can, in fact, be derived directly from the tunneling spectrum. However, in the absence of a generally accepted model for the gap function and the DOS in HTSC, such a quantitative analysis is not straightforward. Thus, in order to compare different spectra, we calculate the gap amplitude 2Δ (in m$e$V) as a half spacing between the conductance peaks at ± 2Δ.

Figure 2 shows for comparison three typical spectra with similar magnitudes obtained on a pure, a Ni- and Zn-doped Bi2212 single crystals. They exhibit the characteristic features of typical SC-insulator-SC junctions [11,12]. There is no much difference among these spectra, they look very similar to each other.

By changing the distance between two pieces of a broken crystal by a differential screw, it is possible to obtain a few tunneling spectra in one B-J. Simple statistics of gap measurements *in one sample* cab be informative about the distribution of the gap magnitude, the predominant character of one of the gaps, and some indication of the origin of the gap(s).

Figure 3 shows the distribution of the magnitude of tunneling gap measured on a pure Bi2212 single crystal (A), a Ni-doped Bi2212 sample (B) and Zn-doped Bi2212 single crystal (C). Note that the vertical scales for each frame in Fig. 3 are different. The choice of the presentation of the data shown in Fig. 3 is due to the maximum number of gap measurements *in one sample* for each set of Bi2212 single crystals. It is important to note that the data in Fig. 3 include only the distribution of the gap magnitude of main tunneling peaks and do not include data for the sub-gap which is often observed in tunneling spectra [5].

One can see in Fig. 3 that the distribution of the magnitude of tunneling gap for each case is in a good agreement with Fig. 1. Secondly, in the pure Bi2212 sample, it is obvious that the d-wave SC gap is predominant. In the Bi2212 single crystal doped with the magnetic impurity, Ni, the d-wave gap is still predominant. In fact, there is almost no difference between Figs. 3a and 3b with the exception for the absolute values of tunneling gap. In contrast to the data for the pure and Ni-doped Bi2212 single crystals, shown in Fig. 3, in the Zn-doped sample, the d-wave SC gap is not 'very popular'. Thus, even simple statistics point out on the magnetic origin of the d-wave SC. The nonmagnetic impurity, Zn 'dilutes' the SC with the magnetic pairing ('swiss cheese' model) and doesn't affect much the spinon SC. However, the most striking fact that Zn 'creates' a gap in the middle of the two gaps with the magnitude of about $2\Delta = 60 - 65$ meV. We will discuss this fact in Section 6. So, from the point of view of statistics, we have a good agreement with the MS model.

## 4. MEASUREMENTS ON Ni-Bi2212: $Ni^{2+}$ OXIDATION STATE

In this Section, we present the data which have been introduced already in our previous work [5], but we need them in order to identify other data. Figure 4 shows *I(V)* and *dI/dV(V)* characteristics measured at 15 K on a Ni-doped Bi2212 single crystal. The temperature dependence of the *dI/dV(V)* is presented in Fig. 5. A normal tunneling *I(V)* characteristic is linear outside the gap-structure, $I \sim V$. The *I(V)* characteristic in Fig. 4 is *striking*, it is almost flat outside the gap. The conductance curve almost doesn't have the background.

The spectra look very *striking* but *senseless* without some additional information. When we have compared these data with inelastic neutron scattering (INS) data on YBa$_2$Cu$_3$O$_{7-x}$ (YBCO) [13,14], we have found a good agreement between the two sets of data [5]. So, it became clear that the so-called resonance peak observed in inelastic neutron scattering measurements and the spectrum in Fig. 4, which appear at the same bais positions as a function of hole concentration, have something in common. In framework of SO(5) theory, Demler and Zhang showed that the resonance-peak position observed in YBCO is in quantitative agreement with the condensation energy of YBCO [15].

The explanation is as follows. The distribution of Ni in this sample probably was not uniform. We tested the DOS in the vicinity of a small cluster of Ni$^{2+}$($S = 1$). The spins $S = 1$ of Ni will allow to propagate in it's vicinity only excitations with $S = 1$. Thus, the spectrum in Fig. 4 can only be explained in two ways: it is a triplet (p-wave) SC state, or it is a SC state mediated by spin-waves. Since we know that in Bi2212, the predominant SC is the d-wave SC mediated by spin-waves, we find that the spectrum in Fig. 4 is a spectrum with the pure magnetic origin, *i.e.* there is no contribution of the spinon SC in the spectrum shown in Fig. 4. The spinon SC is absent because the DOS was tested in a place where charged stripes are absent. Tunneling spectra obtained far from impurities look normal (see, for example, the spectrum B in Fig. 2).

Now we discuss temperature dependencies of the SC gaps. Figure 6 shows temperature dependencies of tunneling spectra of Bi2212. The curve A displays a temperature dependence of the spectra shown in Fig. 5, which have the purely magnetic origin. The curve B in Fig. 6 corresponds to the temperature dependence of tunneling gap observed in (0, $\pi$) direction on the Fermi surface [5], *i.e.* it corresponds to the temperature dependence of the combination of the predominant d-wave SC gap and spinon-SC gap (see the inset in Fig. 1). The curve C in Fig. 6 is a *typical* temperature dependence for a maximum tunneling gap or for a gap which is close to the maximum value [12,5], *i. e*. this is a temperature dependence of the spinon-SC gap since the d-wave gap is absent in ($\pi$, $\pi$) direction on the Fermi surface (see the inset in Fig. 1). We need these data in the next Section in order to find out the origin of some abnormal spectra.

## 5. MEASUREMENTS ON Ni-Bi2212: Ni$^{3+}$ OXIDATION STATE

In this Section, we present spectra which, at the first sight, look very abnormal. However, the comparison with the data presented in Section 4 shows that these abnormal spectra most likely corresponds to the Ni$^{3+}$ oxidation state in Bi2212.

Figures 7 and 8 show spectra as a function of temperature obtained on the same Ni-Bi2212 sample as the spectra shown in Figs. 4 and 5. One can suggest that this Ni-Bi2212 single crystal is 'special'. However, similar spectrum was detected in another Ni-Bi2212 single crystal, which is shown in Fig. 9. In order to identify the origin of the spectra in Figs. 7, 8 and 9, we present the temperature dependence of these tunneling spectra in Fig. 10 [16].

Let's analyze the presented spectra and their temperature dependencies. First of all, it is easy to identify the origin of the smallest SC gap shown in Fig. 8. It's temperature dependence A in Fig. 10 is very similar to the temperature dependence A shown in Fig. 6, which has the pure magnetic origin. The magnitude of the Josephson current in the spectrum in Fig. 8 is much higher than in the other spectra shown in Figs. 7 and 9, and looks similar to the magnitude of the Josephson current in the spectrum of the d-wave SC gap in Fig. 4. The magnitudes of the two gaps shown in Figs. 4 and 8 are approximately equal. Consequently, the smallest SC gap shown in Fig. 8 is the d-wave gap.

The magnitudes of the gaps corresponding to the narrow tunneling peaks shown in Figs. 7, 8 and 9 are around the magnitude of the spinon-SC gap at $p = 0.11$ (see Fig. 1). However, their temperature dependencies in Fig. 10 are similar to the temperature dependence B shown in Fig. 6, which corresponds to the temperature dependence of the combination of the spinon and d-wave SC gaps and different from the temperature dependence of the spinon-SC gap (curve C in Fig. 6). The values of the Josephson current in the spectra shown in Figs. 7 and 9 are small, however non-zero. As we remember, the Josephson current is a characteristic of the d-wave SC in Bi2212. So, we have a puzzling situation: the magnitudes of the SC gaps corresponding to the narrow peaks in the tunneling spectra shown in Figs. 7, 8 and 9 correspond to the magnitude of the spinon-SC gap at $p = 0.11$ and, on the other hand, there is an involvement of the SC with the magnetic origin. In framework of the MS model, there is only one explanation of this experimental fact. These narrow tunneling peaks are due to the $Ni^{3+}(S = 1/2)$ oxidation state in Bi2212, which occurs on charged stripes. This means that $Ni^{3+}$ ions (probably, clusters) participate in the two SCs at the same time! We discuss the data in Section 7.

One may wonder (i) why the spectra shown in Figs. 7, 8 and 9 are asymmetrical relatively zero bias inside the main tunneling peaks, and (ii) why there are so many small peaks inside of the main tunneling peaks (see, for example, Fig. 7). (i) We know that the d-wave SC pairing occurs due to magnons. Spin-waves excited by applied dc current in Cu/Co multilayers are asymmetrical with the direction of the dc current [17]. It may explain the asymmetry of our tunneling spectra. (ii) The shape of the small tunneling peaks inside the main peaks in the spectrum shown in Fig. 7 looks very similar to the shape of the tunneling spectrum of a magnon in an AF medium [6]. In both cases, there is a small step before the peak. This means that the small peaks inside the main tunneling peaks shown in Figs. 7, 8 and 9 correspond to the excitation of magnons in AF regions. However, there is also a difference between these two cases, namely, in bias scale. The small peaks in Fig. 7 are smaller in a scale than the peak in the tunneling spectrum of a magnon [6]. This difference in a scale can be explained by the fact that the process of the magnon excitations in the two cases happens in two different environments. The magnon excitation in an AF medium observed in Ref. 6 occurs in the normal-state environment. In our case, magnons are excited in the SC state. This fact explains the difference in bias scale between the tunneling spectra corresponding to the excitation of magnons.

Let's now discuss measurements on Zn-doped Bi2212 single crystals and, then, analyze all data in Section 7.

## 6. MEASUREMENTS ON Zn-Bi2212

In Section 3, simple statistics show that the nonmagnetic impurity, Zn 'dilutes' the SC with the magnetic pairing ('swiss cheese' model), and, it seems that it doesn't affect much the spinon SC. However, Zn 'creates' a gap with the magnitude of about 60 - 65 meV. Figure 11 shows a spectrum obtained on a Zn-Bi2212 single crystal. One can clearly see in Fig. 11 three different gaps. The peaks at ±105 mV and ±40 mV correspond to the spinon and d-wave SC gaps, respectively. One can see that there is a gap with the value of $2\Delta = 65$ meV.

Let's discuss the origin of the gap with $2\Delta = 65$ meV, which most likely occurs due to the presence of $Zn^{2+}(S = 0)$. There are two possible origins of this gap, namely, it can be either a polaron or spinon gap. We know that the $T_c$ value is defined by the d-wave SC. The increase of the magnitude of the d-wave gap would imply the increase of $T_c$. However, it is not the case. Moreover, Zn is well known as a particularly harmful impurity for $T_c$ in cuprates. So, it is most likely that the gap with $2\Delta = 65$ meV is a partially damaged spinon-SC gap.

There are two independent ways to explain the appearance of the spinon gap with $2\Delta = 65$ meV. (i) According to the Anderson's theorem, a nonmagnetic impurity will not affect the DOS with an isotropic OP. If the spinon-SC OP has a (s+d) mixed symmetry, then, Zn affects locally only a d-component keeping a s-wave component unchanged. Thus, the gap with $2\Delta = 65$ meV is only a part of the complete spinon-SC gap. Then, in framework of the MS model, this implies that a $Zn^{2+}$ creates 'voids' with the diameter larger than the average distance between charged stripes because the spinon SC occurs on charged stripes and it is the only way for a Zn to affect the spinon-SC gap. (ii) The second explanation is as follows. The spinon OP may have either a s-wave or (s+d) mixed symmetry. A $Zn^{2+}$ affects locally the spin-gap which occurs due to AF correlations [2]. The spinon-SC gap depends completely on the spin-gap and will be damaged if the spin-gap is partially destroyed. Consequently, the gap with $2\Delta = 65$ meV is the spinon-SC gap which occurs in regions with the weakened spin-gap. In this case, it is not necessary for Zn to create voids with the diameter larger than the average distance between charged stripes. The diameter of voids can be smaller than the distance between charged stripes. In fact, both presented explanations may be true.

## 7. DISCUSSION: THE MCS MODEL

Most of the effects of impurities substituted for Cu in Bi2212 have been discussed in our previous work [5]. However, the $Ni^{3+}(S = 1/2)$ oxidation state and it's effect on the DOS in Bi2212 has been not considered before. In Section 5, we have found that a $Ni^{3+}(S = 1/2)$ may participate simultaneously in the two SCs. In fact, it is very important result. The $Ni^{3+}$ participation in the two SCs simultaneously implies that the two SCs exist in different degrees of freedom, *i.e.* in different direction. In the MS model, they coexist 'in parallel' and it is impossible for a $Ni^{3+}$ ion to participate in the two SCs at the same time. The $CuO_2$ planes are two-dimensional. The only possible solution of this problem, if we consider that the spinon SC occurs along charged stripes *and* the SC with the magnetic pairing occurs in perpendicular direction to stripes. So, we have to modify the MS model.

Here, we present a Magnetic Coupling of Stripes (MCS) model of the HTSC in Bi2212, which is, in fact, the *modified* MS model. The scenario in framework of the MCS model is very similar to a scenario described by Emery with co-workers [2]. The main difference between the two models is that the coherent SC state of the spinon SC is established differently in the two models. In framework of the MCS model, the coherent state is established by magnons and not by the Josephson coupling as described in Ref. 2. Magnons are excited in AF domains of $CuO_2$ planes by motions of charged stripes. We know that stripes fluctuate [3]. An electron moving into AF medium excites spin-waves [6]. So, fluctuating charged stripes excite magnons. That means that, in the MCS model, a magnetic polaron (in the MS model) is replaced by a charged stripe which has a virtual spin polarization around it. There is only one SC, the spinon SC, but there are two mechanisms, one for pairing and one for the coherent state. In BCS theory, electrons couple to each other by phonons. The coherent state among the Cooper pairs is established also by phonons. In hole-doped cuprates, there exists one SC, but the pairing and coherent-state mechanisms are different: spinons couple to each other due to the existence of the large spin-gap in order to low their total free energy, *and* the coherent state of the spinon SC is established by magnons. Charged stripes with the spinon SC couple to each other by spin-waves which are excited due to its' motions. Impurities pin charged stripes [3], consequently, reduce the magnetic interstripe coupling. As a consequence, the $T_c$ value is reduced too. Thus, there is one type of carriers but they exhibit different properties in different directions: *fermionic* along charged stripes and *polaronic* perpendicular to stripes.

Further, we discuss shortly the SC mechanism in other cuprates. The MCS scenario is described for hole-doped cuprates with the high value of $T_c$, *i. e.* for cuprates with more than one $CuO_2$ plane. Phonons do not play important role in the magnetic coupling of stripes since electron-magnon interactions are much stronger than electron-phonon interactions in AF compounds [6].

However, in $La_{2-x}Sr_xCuO_4$ (LSCO) with the single $CuO_2$ plane, magnetic interactions are weaker than, for example, in Bi2212. So, charged stripes carrying the spinon SC couple to each other not only by magnons but also by phonons [18,19]. In the electron doped $Nd_{2-x}Ce_xCuO_4$ (NCCO) cuprate, it seems that, for some reasons, magnons can not propagate, so, stripes use only phonons for coupling *or/and* the Josephson coupling. Consequently, the MCS model has to be named as a Magnetic-Phonon Coupling of Stripes (MPCS) model for LSCO and as a Phonon Coupling of Stripes (PCS) model for NCCO (or as a Josephson Coupling of Stripes (JCS)). The difference in the mechanism of the interstripe coupling in different cuprates *explains* the difference in the symmetry of the predominant OP in different cuprates.

Let's analyze shortly the other data. We have to admit that impurities help to understand the mechanism of the HTSC. What did we found?

(i) We have seen that the Ni doping causes the appearance of two abnormalities in the quasiparticle DOS in Bi2212. The Zn doping creates only one effect in the quasiparticle DOS. It is in a good agreement with the number of possible oxidation states for each element. From our study along, we are not able to conclude what Ni oxidation state is preferable in Bi2212: $Ni^{2+}$ or $Ni^{3+}$. It seems that both states are present in Bi2212.

(ii) Our study shows that Fig. 1 corresponds to the reality. Deutscher [20] has presented strong indications in support of the validity of Fig. 1. Although, the spinon-SC gap is an excitation gap in his interpretation. Recently, it has been shown that a tunneling gap with the maximum magnitude if a SC gap [12].

In the new interpretation, Fig. 1 shows the maximum magnitudes of the two gaps in Bi2212: the spinon gap along stripes and the d-wave gap of the interstripe coupling by magnons. The $2\Delta/k_BT_c$ value for the d-wave gap in Bi2212 is around 5.3. The $E_r/k_BT_c$ value for YBCO is of the order of 5.1 - 5.7, where $E_r$ is the energy of the resonance peak in INS measurements [13,14]. For the magnetic pairing, the $E_r$ is simply equal to $2\Delta$. For the spinon SC along charged stripes, the question still remains: what happens at points $p = 0.05$ and $p = 0.27$? Does the spinon SC exist on charged stripes below $p = 0.05$ and above $p = 0.27$?

(iii) In Introduction, we have mentioned about a sub-gap in tunneling spectra, which was associated with the g-wave SC with the magnetic pairing. The sub-gap can bee also seen in Fig. 8. In Zn-Bi2212 single crystals, we did not observed the sub-gap. Theoretically, the g-wave SC is more robust with the change of hole concentration than the d-wave SC [7]. This is in contradiction what we observe. Indeed, more research is needed to find out the answer about the origin of the sub-gap in tunneling spectra. But it is absolutely clear that the sub-gap has the magnetic origin.

Finally, frequently asked question is why ARPES data [21] are so different from tunneling measurements [22]. It is difficult to find a direct answer to this question. We know that the ARPES energy resolution (~15 meV) is, at least, one order of magnitude worse than in tunneling measurements. The escape depth is only 3 Å from the sample surface. Typically, a time of one

APRES measurement is a few hours while it is a matter of seconds in tunneling measurements. The ARPES data provide more direct information for the valence-charge distribution. So, APRES measurements are not sensitive to the pure spin channel. These are the main differences between APRES and tunneling measurements.

## 8. CONCLUSION

We summarize the content of present work. We presented direct measurements of the density-of-state by tunneling spectroscopy on pure, Ni- and Zn-doped $Bi_2Sr_2CaCu_2O_{8+x}$ (Bi2212) single crystals at low temperatures using the break-junction technique. We used Ni and Zn impurities doped into Bi2212 to test the MS model. We found that Ni doped into Bi2212 can be in two oxidation states: $Ni^{2+}(S=1)$ and $Ni^{3+}(S=1/2)$. A Zn is always in the $Zn^{2+}(S=0)$ oxidation state. We show that magnetic (Ni) and nonmagnetic (Zn) impurities doped into $CuO_2$ planes affect both the $T_c$ and the density of states in Bi2212 but do it differently. In order to explain the data, we find that the MCS model can describe the superconductivity in Bi2212 better than the MS model. The MCS model is a *modification* of the MS model. In the MCS model, there is only one superconductivity – the spinon superconductivity along charged stripes. The coherent state of the spinon superconductivity is established by spin-waves which are excited by motions of charged stripes in antifferomagnetic domains of $CuO_2$ planes. Thus, in the MCS model, the superconductivity has the two mechanisms: along charged stripes and perpendicular to stripes. Consequently, carriers exhibit different properties in different directions: *fermionic* along charged stripes and *polaronic* perpendicular to stripes.

I thank R. Deltour for support, D. Ciurchea for performing EDAX examination. This work is supported by PAI 4/10.

**FIGURE CAPTIONS**

FIG. 1. Idealized phase diagram for two SC gaps in Bi2212: straight line (spinon gap) and parabolic line (d-wave gap). Dots show gap magnitudes for two cases: $p = 0.19$ and $p = 0.11$. Inset: shapes of two SC gaps on the Fermi surface in overdoped Bi2212: black area (d-wave gap) and outlined area (spinon gap having either a s-wave or mixed (s+d) symmetry). The shapes of two gaps are shown schematically.

FIG. 2. Tunneling spectra obtained on (a) a pure Bi2212 single crystal; (b) a Ni-doped Bi2212 single crystal, and (c) a Zn-doped Bi2212 single crystal. The conductance scale corresponds to the spectrum A; the spectra B and C are shifted for clarity by 3 and 6 units, respectively. The spectra are normalized at -200 mV.

FIG. 3. Statistics of the magnitude of tunneling gap measured in one Bi2212 single crystal: (a) pure; (b) Ni-doped, and (c) Zn-doped. Pay attention to the difference in the vertical scale for each frame.

FIG. 4. *I(V)* and *dI/dV(V)* characteristics measured at 15 K on a Ni-doped Bi2212 single crystal with $T_c$ = 75 K. The conductance curve is normalized at -200 meV.

FIG. 5. Tunneling spectra vs. temperature for the Ni-doped Bi2212 single crystal (see Fig. 4). The conductance scale corresponds to the 70.3 K spectrum, the other spectra are offset vertically for clarity. The conductance curves are normalized at -150 meV (or nearest point).

FIG. 6. Temperature dependencies of the quasiparticle DOS in Bi2212 single crystals: A (the SC gap of the pure magnetic origin, shown in Fig. 5); B (a SC gap measured at $(0, \pi)$ on the Fermi surface), and C (a SC gap measured in $(\pi, \pi)$ direction on the Fermi surface) [5].

FIG. 7. Tunneling spectra vs. temperature for the Ni-doped Bi2212 single crystal (see Fig. 4). The conductance scale corresponds to the 72.8 K spectrum, the other spectra are offset vertically for clarity. The conductance curves are normalized at -150 meV (or nearest point).

FIG. 8. Tunneling spectra vs. temperature for the Ni-doped Bi2212 single crystal (see Fig. 4). The conductance scale corresponds to the 80 K spectrum, the other spectra are offset vertically for clarity. The conductance curves are normalized at -130 meV (or nearest point).

FIG. 9. Tunneling spectra vs. temperature for a Ni-doped Bi2212 single crystal. The conductance scale corresponds to the 72.3 K spectrum, the other spectra are offset vertically for clarity. The conductance curves are normalized at -150 meV (or nearest point).

FIG. 10. Temperature dependencies of the quasiparticle DOS in Ni-doped Bi2212 single crystals: square (tunneling spectra shown in Fig. 7); diamonds and dots (tunneling spectra shown in Fig. 8), and triangles (tunneling spectra shown in Fig. 9). The dependence A (dots) corresponds to the smallest SC gap in Fig. 8.

FIG. 11. Normalized tunneling spectrum obtained on a Zn-doped Bi2212 single crystal.

**(NOTE: the figures are in .ps format with poor quality for graphs)**

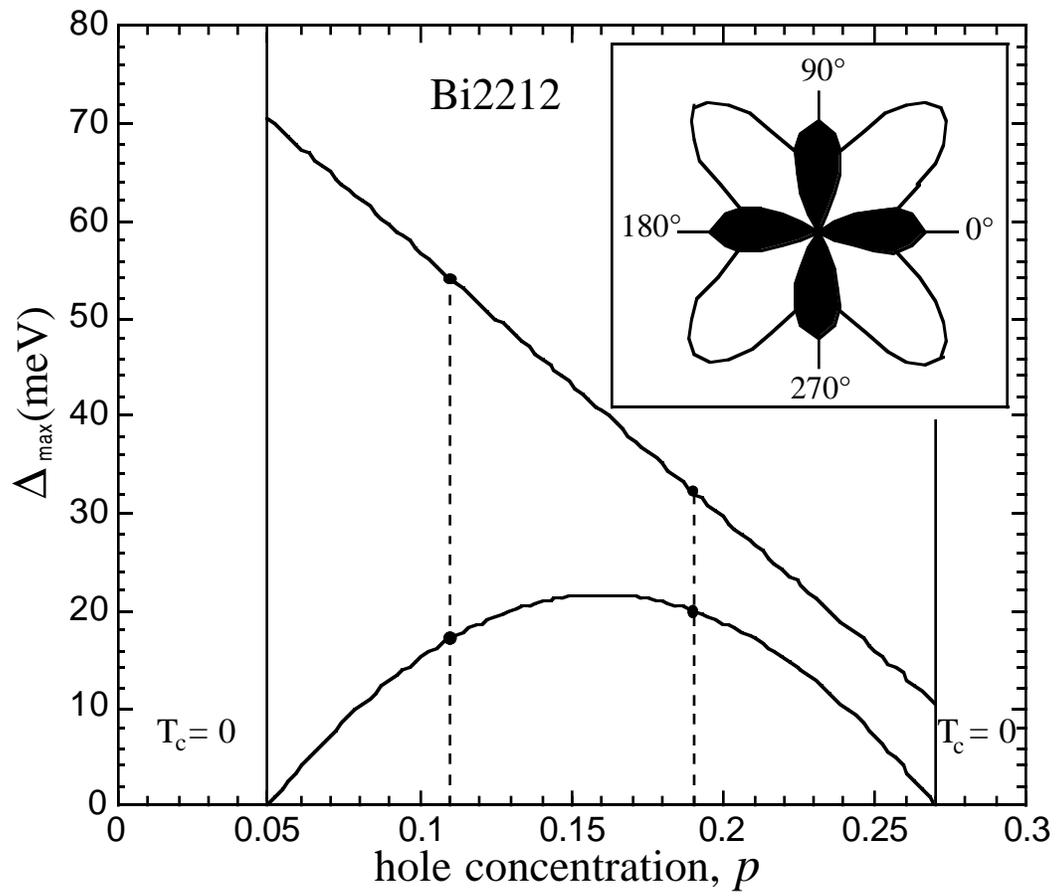

FIG. 1

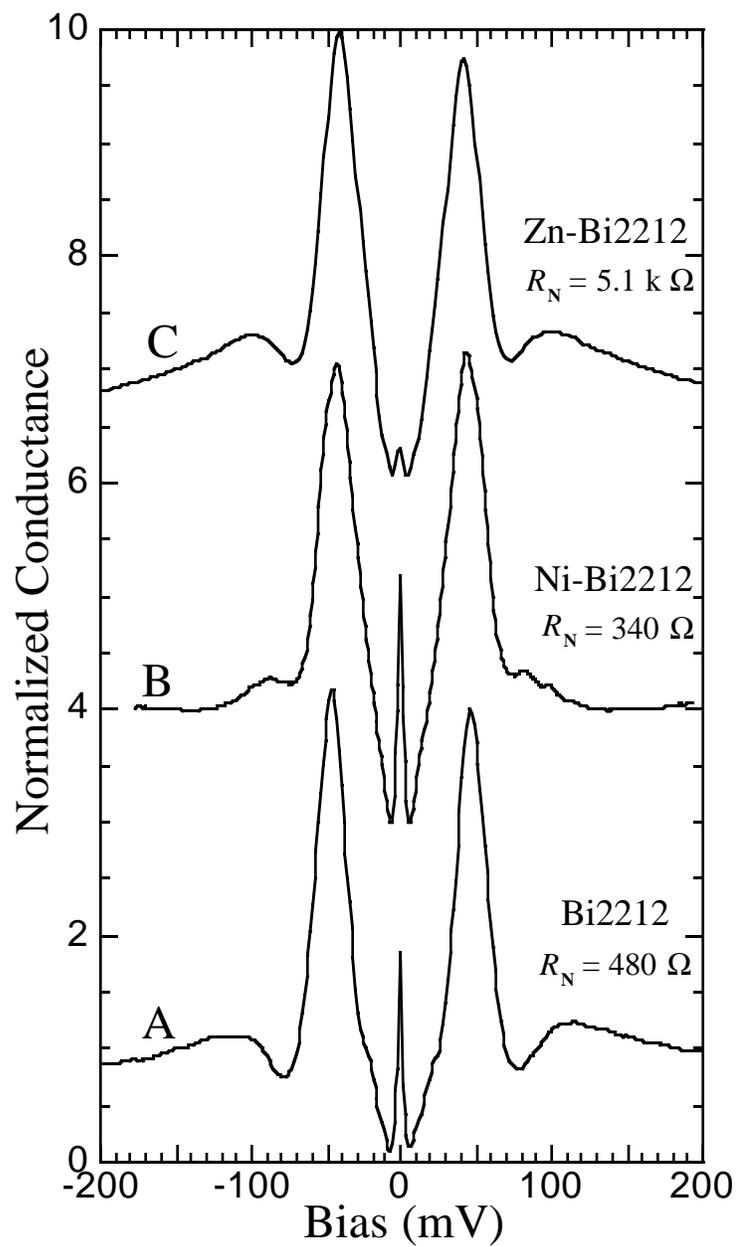

FIG. 2

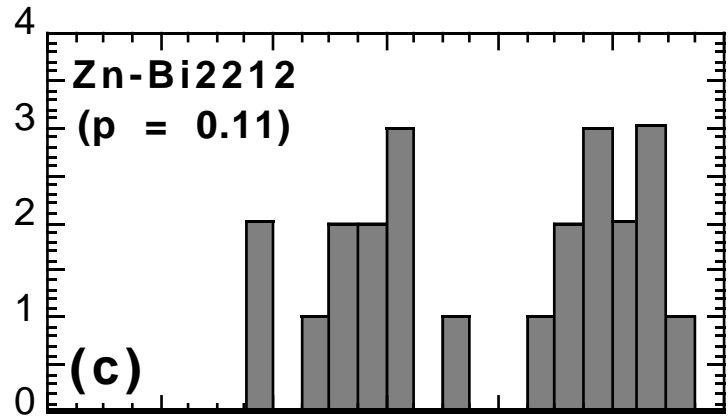
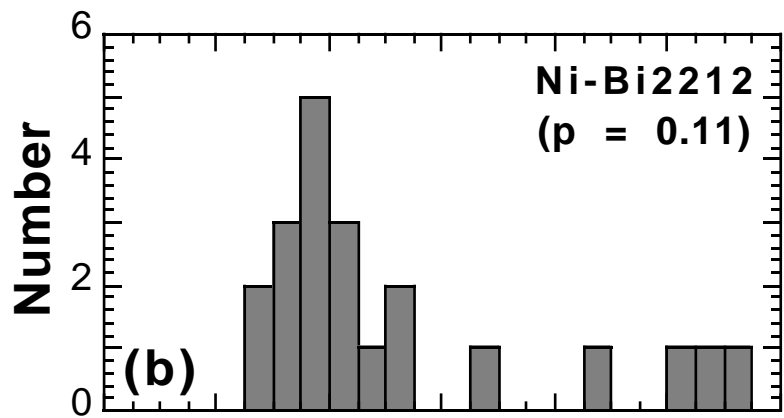

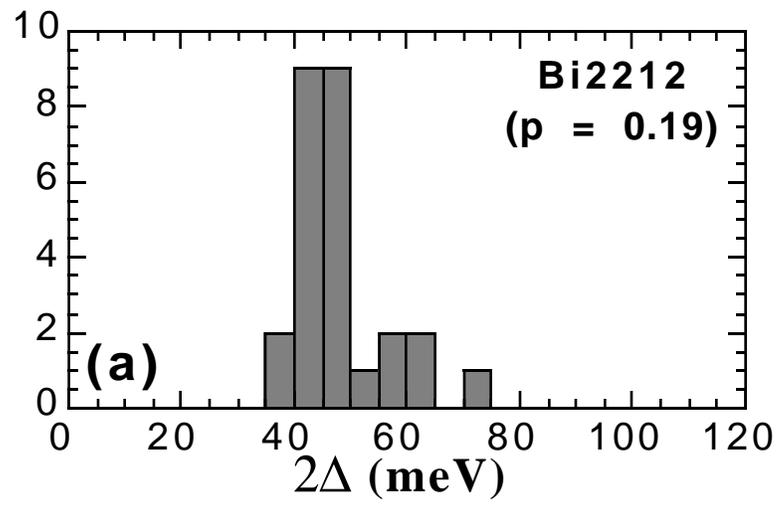

FIG. 3

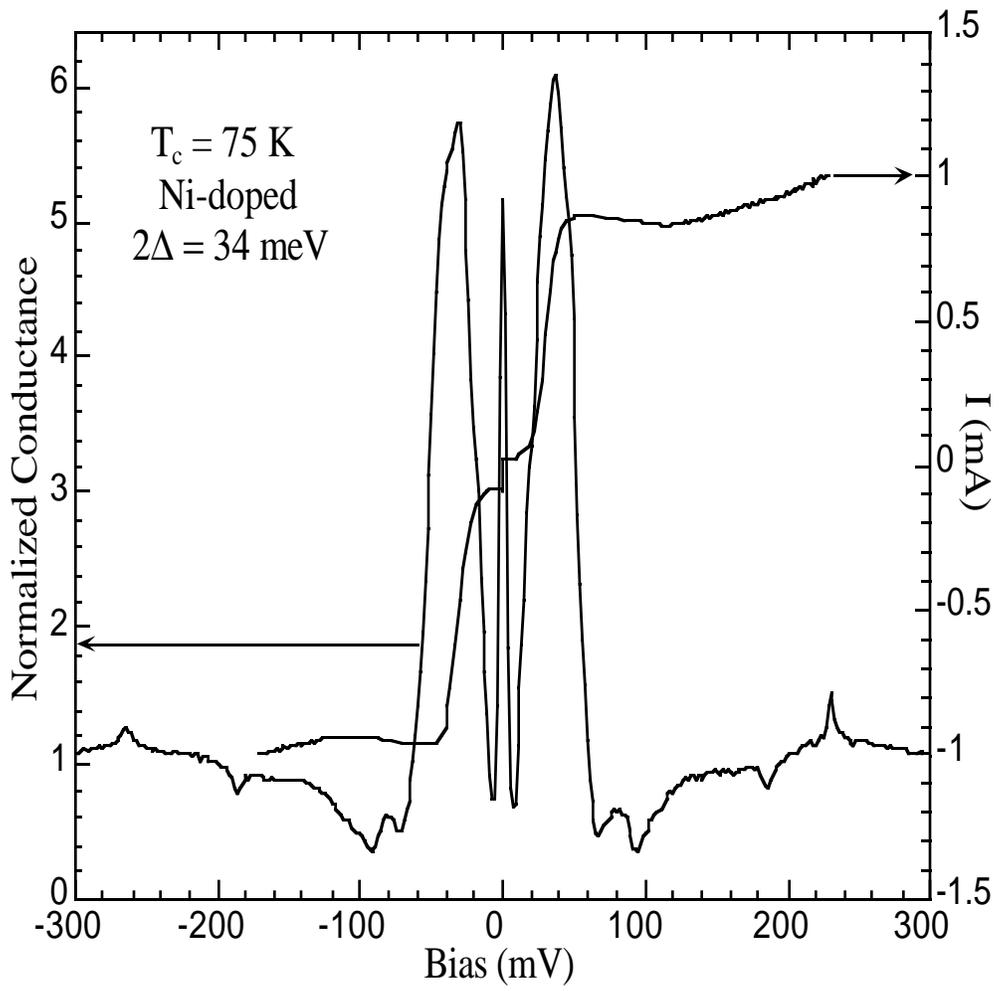

FIG. 4

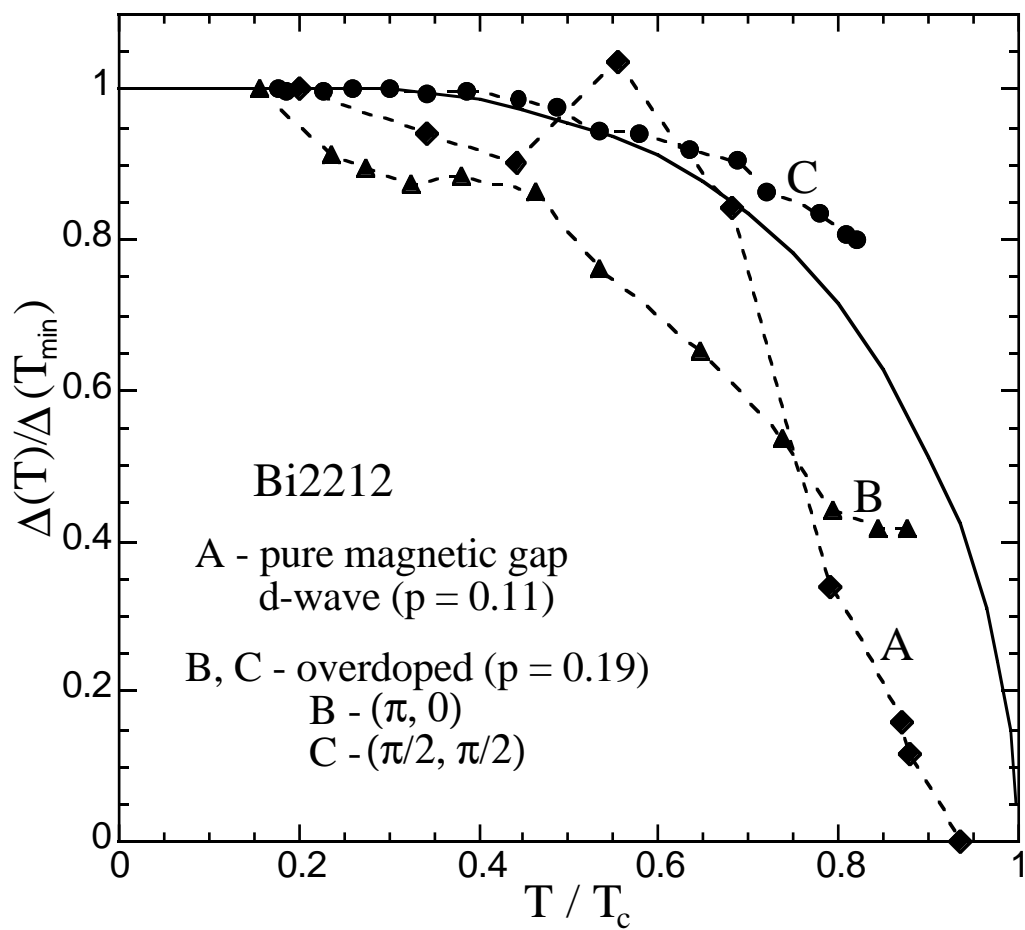

FIG. 5

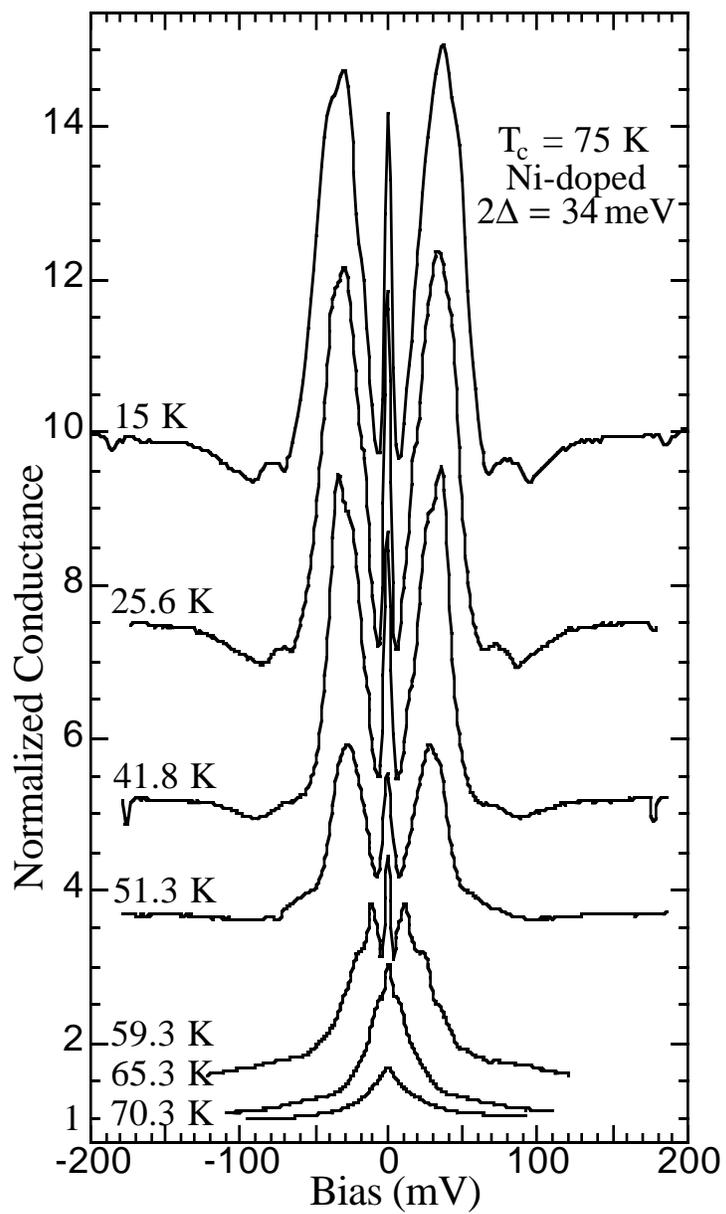

FIG. 6

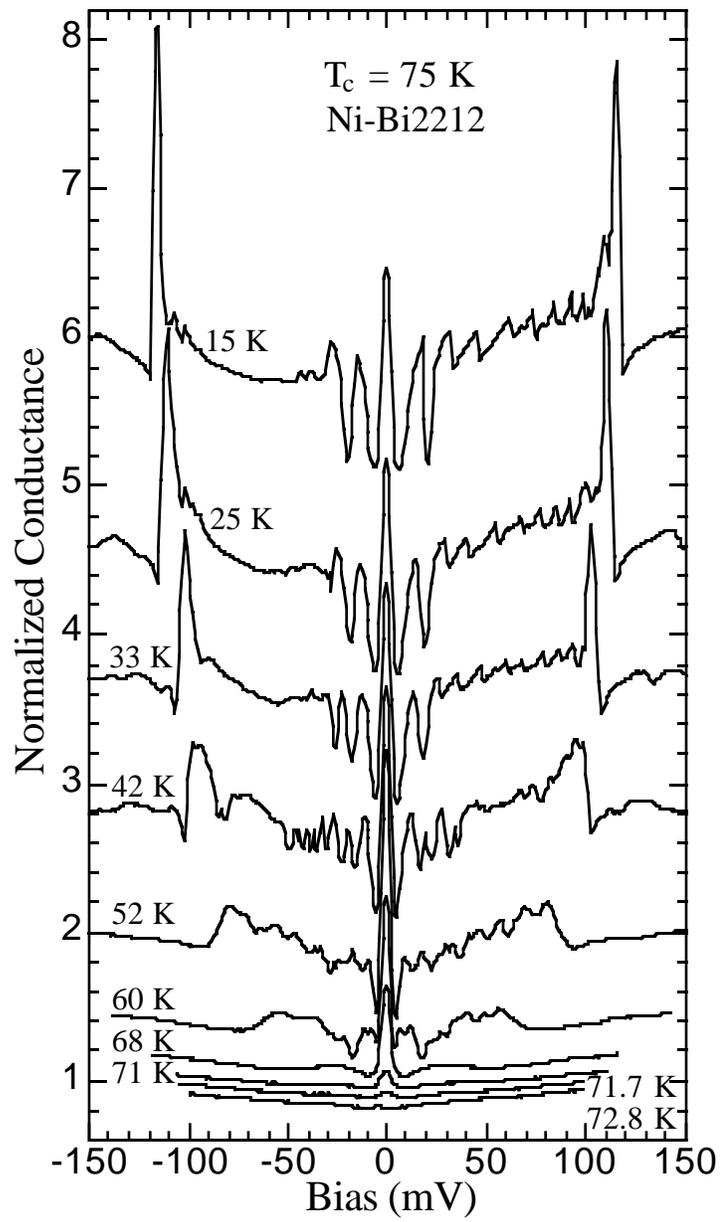

FIG. 7

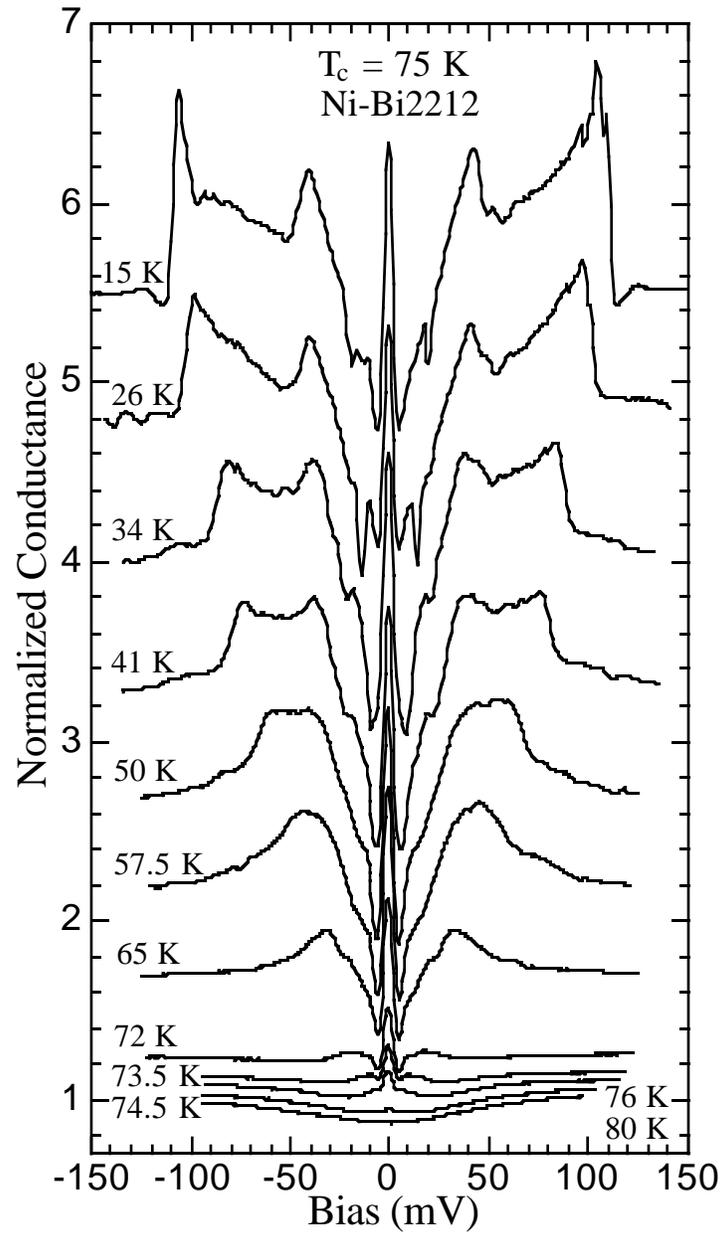

FIG. 8

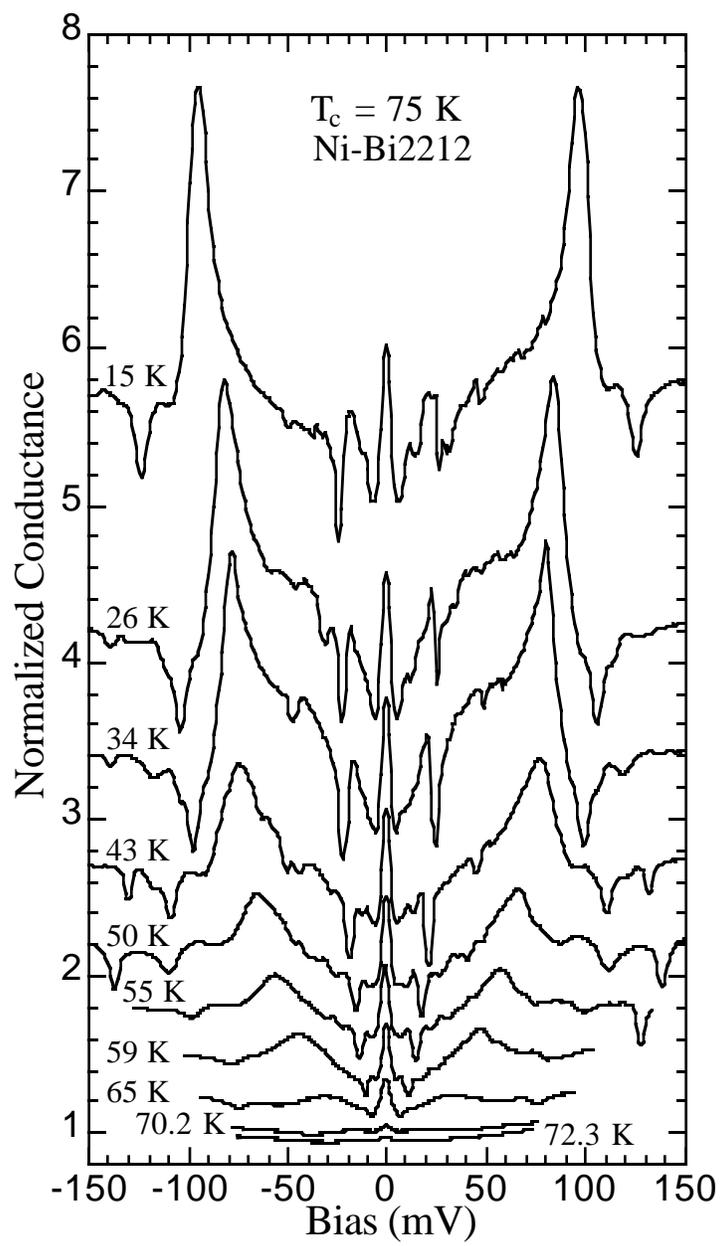

FIG. 9

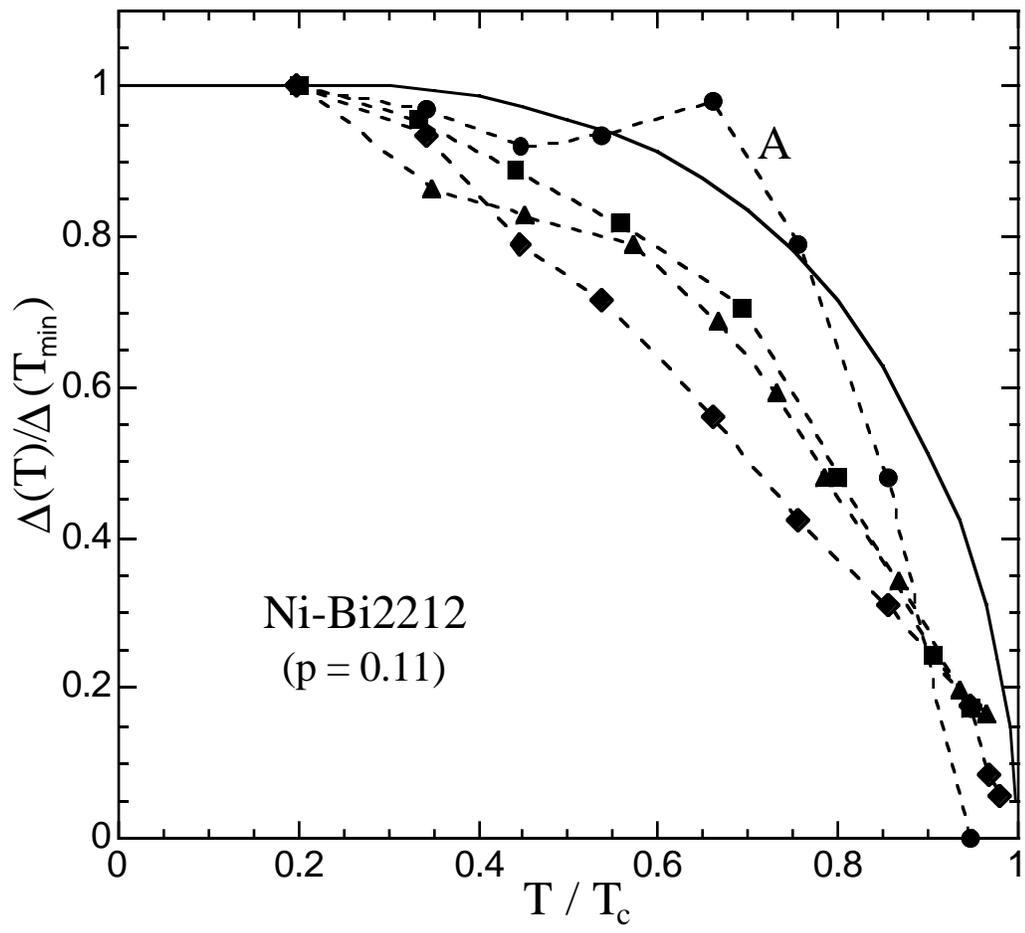

FIG. 10

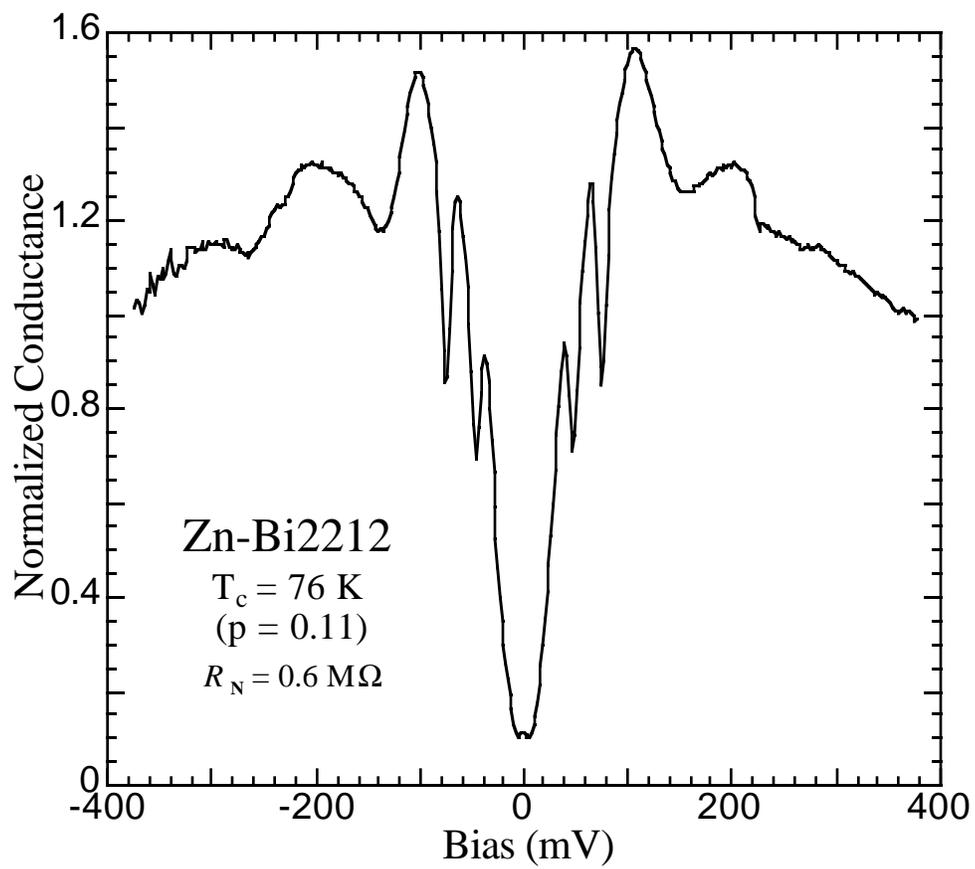

FIG. 11